**Second-Order, Biconformally Invariant  
Scalar-Tensor Field Theories  
in a Four-Dimensional Space**

by

Gregory W. Horndeski  
Adjunct Associate Professor of Applied Mathematics  
University of Waterloo  
Waterloo, Ontario  
Canada  
N2L 3G1

email:  
ghorndeski@uwaterloo.ca  
or  
horndeskimath@gmail.com

October 9, 2022

**ABSTRACT**

In this paper I shall consider field theories in a space of four-dimensions which have field variables consisting of the components of a metric tensor and scalar field. The field equations of these scalar-tensor theories will be derived from a variational principle using a Lagrange scalar density which is a concomitant of the field variables and their derivatives of arbitrary, but finite, order. I shall consider biconformal transformations of the field variables, which are conformal transformations which affect both the metric tensor and the scalar field. A necessary and sufficient condition will be developed to determine when the Euler-Lagrange tensor densities are biconformally invariant. This condition will be employed to construct all of the second-order biconformally invariant scalar-tensor field theories in a space of four-dimensions. It turns out that the field equations of these theories can be derived from a linear combination of (at most) two second-order Lagrangians, with the coefficients in that linear combination being real constants.

## SECTION 1: INTRODUCTION

In June, 2022 Prof. Paul J.Steinhardt contacted me with a problem. He wanted to know what all of the conformally invariant second-order scalar-tensor field theories were, since that would assist him and his students in their endeavors to construct models of bouncing universes. Well, in Ref. 1, I constructed all such field theories of arbitrary differential order, and I directed him to that result. His response indicated that he was not interested in conformal transformations that only affected the metric, but wanted the transformation to also affect the scalar field. I did not know the answer to that question, and the purpose of this paper is to provide the answer when the field tensor densities are of second-order.

To begin, let me introduce some terminology.

When I refer to a theory as a scalar-tensor field theory that means that the field variables are the local components of a pseudo-Riemannian metric tensor, $g_{ab}$, and a scalar field $\varphi$, and that the field equations are derivable from a Lagrangian which is a scalar density tensorial concomitant of the form

$$L = L(g_{ab}; g_{ab,c}; \ldots ; \varphi; \varphi_{,c}; \ldots ) \qquad \text{Eq.1.1}$$

which is of finite differential order and a comma is used to denote partial differentiation with respect to the local coordinates employed to compute the metric tensor's components. The associated Euler-Lagrange tensor densities are defined by

$$E^{ab}(L) \equiv \frac{\delta L}{\delta g_{ab}} := \frac{\partial L}{\partial g_{ab}} - \frac{d}{dx^c} \frac{\partial L}{\partial g_{ab,c}} + \ldots \qquad \text{Eq.1.2a}$$

$$E(L) \equiv \frac{\delta L}{\delta \varphi} := \frac{\partial L}{\partial \varphi} - \frac{d}{dx^c} \frac{\partial L}{\partial \varphi_{,c}} + \ldots \qquad \text{Eq.1.2b}$$

In Ref. 2, I show that in a space of any dimension the Euler-Lagrange tensor densities are not independent but are related by the identity

$$E_a{}^b(L)_{|b} = \tfrac{1}{2}\varphi_{,a}E(L) \qquad \text{Eq.1.3}$$

where the vertical bar denotes covariant differentiation with respect to the Levi-Civita connection determined by the metric tensor, and repeated indices are summed. The definitions of $E^{ab}(L)$ and $E(L)$ given above differ from those I used in Refs. 2 and 3, by a minus sign, but that will not prove to be an obstacle in what is done here. In addition the geometric quantities that I shall use here are defined as they were in Refs. 2 and 3, with $V^i{}_{|jk} - V^i{}_{|kj} := V^h R_h{}^i{}_{jk}$, where $V^i$ is an arbitrary vector field, $R_{hj}:=R_h{}^i{}_{ji}$ and $R:= g^{hj}R_{hj}$. I shall also let $\varphi_a:=\varphi_{,a}$; $\varphi_{ab}:=\varphi_{|ab}$, $\Box\varphi:= g^{ab}\varphi_{ab}$ and $g:= |\det(g_{ab})|$.

For scalar-tensor field theories I define a biconformal transformation of degree k ($k\in\mathbb{R}$) by:

$$g_{ab}\to g'_{ab} := e^{2\sigma}g_{ab} \quad \text{and} \quad \varphi \to \varphi' := e^{k\sigma}\varphi \qquad \text{Eq.1.4}$$

where $\sigma$ is an arbitrary scalar field.

If L is a Lagrangian of the form given in Eq.1.1, then the transformed Lagrangian L' is obtained from L by simply replacing $g_{ab}$, $\varphi$ and their derivatives by $g'_{ab}$, $\varphi'$ and their derivatives. L is said to be conformlly invariant if

$$L'(g'_{ab}; g'_{ab,c}; \ldots ; \varphi'; \varphi'_{,c}; \ldots) = L(g_{ab}; g_{ab,c}; \ldots ; \varphi; \varphi_{,c}; \ldots) . \qquad \text{Eq.1.5}$$

This means that if you take L as in Eq.1.1 then

$$L(e^{2\sigma}g_{ab}; (e^{2\sigma}g_{ab})_{,c}; \ldots ; e^{k\sigma}\varphi; (e^{k\sigma}\varphi)_{,c}; \ldots) = L(g_{ab}; g_{ab,c}; \ldots ; \varphi; \varphi_{,c}; \ldots) . \qquad \text{Eq.1.6}$$

When I say that a Lagrangian L of the form given in Eq.1.1 is biconformally invariant up to a divergence I mean that there exists a vector density concomitant $V^i$ built from $g_{ab}$, $\varphi$, $\sigma$ and their derivatives which is such that Eq.1.5 is replaced by

$$L' = L + \frac{d V^i}{dx^i} . \qquad \text{Eq.1.7}$$

A well-known property of the Euler-Lagrange operator is that it annihilates divergences. (For a direct proof of this fact see Lovelock [4].) In this case we can use Eq.1.7 to write

$$\frac{\delta L'}{\delta g'_{ab}} = \frac{\delta}{\delta g'_{ab}}(L + \frac{d V^i}{dx^i}) = \frac{\delta}{\delta g_{rs}}(L + \frac{d V^i}{dx^i}) \frac{\delta g_{rs}}{\delta g'_{ab}} \ . \qquad \text{Eq.1.8}$$

Since

$$\frac{\delta g_{rs}}{\delta g'_{ab}} = \frac{\partial g_{rs}}{\partial g'_{ab}} = \tfrac{1}{2} e^{-2\sigma}(\delta^a_r \delta^b_s + \delta^a_s \delta^b_r)$$

(for those unfamiliar with the process of differentiating tensor concomitants please see Appendix A in [1]) Eq.1.8 tells us that

$$E^{ab}(L)' \equiv \frac{\delta L'}{\delta g'_{ab}} = e^{-2\sigma} \frac{\delta L}{\delta g_{ab}} = e^{-2\sigma} E^{ab}(L) \ . \qquad \text{Eq.1.9}$$

Similarly, Eq.1.7 tells us that when L is biconformally invariant up to a divergence then

$$E(L)' \equiv \frac{\delta L'}{\delta \varphi'} = e^{-k\sigma} \frac{\delta L}{\delta \varphi} = e^{-k\sigma} E(L) \ . \qquad \text{Eq.1.10}$$

Eqs. 1.9 and 1.10 allow us to deduce that when a scalar-tensor field theory is such that its Lagrangain is biconformally invariant of degree k up to a divergence then

$$E^a_{\ b}(L)' = g'_{bc} E^{ac}(L)' = g_{bc} E^{ac}(L) = E^a_{\ b}(L) \qquad \text{Eq.1.11}$$

and

$$\varphi' E(L)' = \varphi E(L) \ . \qquad \text{Eq.1.12}$$

When I say that the field tensor densities of a scalar-tensor field theory are biconformally invariant I mean that they satisfy Eqs. 1.11 and 1.12.

We require an easy way to determine when the field tensor densities of a scalar-tensor field theory are biconformally invariant of degree k. This is provided by the following:

**Proposition 1**: Let $E^{ab}(L)$ and $E(L)$ be the Euler-Lagrange tensor densities of a scalar-tensor field theory. This theory will be biconformally invariant of degree k if and only if

$$2 g_{ab} E^{ab}(L) + k \varphi E(L) = 0 \ . \qquad \text{Eq.1.13}$$

If $E^{ab}(L)$ and $E(L)$ satisfy Eq.1.13 then L is biconformally invariant of degree k up to a divergence.

**Proof:** The proof is modeled on the proof of **Proposition 2.1** in Ref. 1.

$\Rightarrow$ The Euler-Lagrange tensor densities $E_a^{\ b}(L)'$ and $E(L)'$ satisfy the primed version of Eq.1.3; *viz.,*

$$E_a{}^b(L)'_{|b} = \tfrac{1}{2} \varphi'_{,a} E(L)' , \qquad \text{Eq.1.14}$$

where "$_{|b}$" denotes covariant differentiation with respect to $g'_{ab}$. Since (see, *e.g.,* Eq.28.3 in Eisenhart, Ref. 5)

$$\Gamma'{}^r{}_{st} = \Gamma^r{}_{st} + \sigma_{,s}\delta^r{}_t + \sigma_{,t}\delta^r{}_s - g_{st} g^{rp} \sigma_{,p} , \qquad \text{Eq.1.15}$$

we easily find that since $E_a{}^b(L)' = E_{ab}(L)$,

$$E_a{}^b(L)'_{|b} = E_a{}^b(L)_{|b} - E_b{}^b(L)\sigma_{,a} . \qquad \text{Eq.1.16}$$

Eqs. 1.3, 1.12, 1.14 and 1.16 combine to give us

$$\tfrac{1}{2}\varphi'_{,a}\varphi\varphi'^{-1}E(L) = \tfrac{1}{2}\varphi_{,a}E(L) - E_b{}^b(L)\sigma_{,a} \qquad \text{Eq.1.17}$$

and since

$$\varphi\varphi'_{,a}\varphi'^{-1} = k\varphi\sigma_{,a} + \varphi_{,a}$$

Eq.1.17 tells us that

$$\tfrac{1}{2}k\varphi\, E(L) = -E_b{}^b(L)$$

which is what we are trying to demonstrate.

$\Leftarrow$ To prove the converse I define a one parameter family of variations from $g_{ab}$, $\varphi$ to $g'_{ab}$, $\varphi'$ by

$$g(t)_{ab} := e^{2\sigma t}g_{ab} \text{ and } \varphi(t) := e^{k\sigma t}\varphi , \ 0\leq t\leq 1 . \qquad \text{Eq.1.18}$$

Using $g(t)_{ab}$ and $\varphi(t)$ in the Lagrangian L gives us a one parameter family of Lagrangians defined by

$$L(t) := L(g(t)_{ab}; g(t)_{ab,c}; \ldots ; \varphi(t); \varphi(t)_{,c}; \ldots ) . \qquad \text{Eq.1.19}$$

Note that $L(0) = L$, and $L(1) = L'$. Upon differentiating Eq.1.19 with respect to t we get

$$\frac{dL(t)}{dt} = \frac{\partial L}{\partial g_{ab}}(g(t),\varphi(t)) \frac{dg(t)_{ab}}{dt} + \frac{\partial L}{\partial g_{ab,c}}(g(t),\varphi(t)) \frac{dg(t)_{ab,c}}{dt} + \ldots + \frac{\partial L}{\partial \varphi}(g(t),\varphi(t)) \frac{d\varphi(t)}{dt} +$$

$$+ \frac{\partial L}{\partial \varphi_{,c}}(g(t),\varphi(t)) \frac{d\varphi(t)_{,c}}{dt} + \ldots \qquad \text{Eq.1.20}$$

where, *e.g.,* $\frac{\partial L}{\partial g_{ab}}(g(t),\varphi(t))$ means to differentiate L as defined in Eq.1.1 with respect to $g_{ab}$ and then evaluate that derivative with $g_{ab}$, $\varphi$ and their derivatives replaced by $g(t)_{ab}$, $\varphi(t)$ and their derivatives.

Using an argument similar to the usual argument to rewrite equations such as Eq.1.20 (*see, e.g.,* Lemma 2.1 in Horndeski Ref. 6) it can be shown that

$$\frac{dL(t)}{dt} = E^{ab}(L(t))\frac{dg(t)_{ab}}{dt} + E(L(t))\frac{d\varphi(t)}{dt} + \frac{d}{dx^c} V^c(t) \qquad \text{Eq.1.21}$$

where the general form of $V^c(t)$ can be found in Ref. 6. Combining Eqs.1.18 and 1.21 allows us to deduce that

$$\frac{dL(t)}{dt} = 2\sigma E^{ab}(L(t))g_{ab}(t) + k\sigma E(L(t))\varphi(t) + \frac{d}{dx^c} V^c(t) . \qquad \text{Eq.1.22}$$

However, since Eq.1.13 must hold when evaluated for every pair of metric tensors and scalar fields, it must hold when evaluated for $g(t)_{ab}$ and $\varphi(t)$. Thus Eq.1.22 reduces to

$$\frac{dL(t)}{dt} = \frac{d}{dx^c} V^c(t) .$$

When this is integrated with respect to t from 0 to 1, we see that L' and L differ by a divergence when Eq.1.13 holds, which completes the proof of the proposition.∎

So now we have at our disposal a very simple algebraic way of discerning whether a scalar-tensor field theory is biconformally invariant of degree k when we have the field tensor densities. This tool will be used in the next section to construct the second-order biconformally invariant scalar-tensor field theories of degree 1. Once these are found it is a simple matter to obtain all of the other biconformally invariant second-order scalar tensor field theories of degree $k \neq 0$, as I shall demonstrate shortly. The k=0 case is treated in Ref. 1 where I show that there exists only one such second-order scalar-tensor theory, and it is generated by the Lagrangian

$$L_{C,k=0} := g^{\frac{1}{2}} f(\varphi) X^2,$$

where $X := g^{ab}\varphi_{,a}\varphi_{,b}$ and f can be taken to be any real-valued function of $\varphi$.

The astute reader probably realizes that if you can construct biconformally invariant scalar-tensor field theories of degree 1 then it should be easy to get them for any degree k≠0. This fact is demonstrated in

**Proposition 2:** Let $L=L(g_{ab}; g_{ab,c};...; \varphi; \varphi_{,c};...)$ be the Lagrangian of a biconformally invariant scalar-tensor field theory of degree 1. If $k \in \mathbb{R}\setminus\{0\}$, let $L_k = L_k(g_{ab}; g_{ab,c};...; \varphi; \varphi_{,c};...)$ be the Lagrangian defined by $L_k := L(g_{ab}; g_{ab,c};...; \varphi^{1/k}; (\varphi^{1/k})_{,c};...)$. $L_k$ is the Lagrangian of a biconformally invariant scalar-tensor field theory of degree k.

**Proof:** Due to **Proposition 1** all we need show is that

$$g_{ab}\frac{\delta L_k}{\delta g_{ab}} + k\varphi\frac{\delta L_k}{\delta \varphi} = 0 .$$  Eq.1.23

Due to the definition of $L_k$ and the chain rule for variational derivatives

$$g_{ab}\frac{\delta L_k}{\delta g_{ab}} = g_{ab} \frac{\delta L(g,\varphi^{1/k})}{\delta g_{ab}}$$  Eq.1.24

and

$$k\varphi\frac{\delta L_k}{\delta \varphi} = k\varphi\frac{\delta L(g,\varphi^{1/k})}{\delta \varphi} \frac{\delta \varphi^{1/k}}{\delta \varphi} = \varphi^{1/k}\frac{\delta L(g,\varphi^{1/k})}{\delta \varphi} = \left[\varphi \frac{\delta L}{\delta \varphi}\right](g_{ab},\varphi^{1/k})$$  Eq.1.25

Eqs. 1.24 and 1.25 combine to tell us that

$$g_{ab}\frac{\delta L_k}{\delta g_{ab}} + k\varphi \frac{\delta L_k}{\delta \varphi} = \left[g_{ab}\frac{\delta L}{\delta g_{ab}} + \varphi\frac{\delta L}{\delta \varphi}\right](g,\varphi^{1/k})$$

which vanishes since $L(g_{ab}; g_{ab,c};...; \varphi; \varphi_{,c};...)$ defines a biconformally invariant scalar-tensor field theory of degree 1. Consequently $L_k$ satisfies Eq.1.23, and hence is biconformally invariant of degree k.∎

So due to **Proposition 2** when searching for biconformally invariant scalar-tensor field theories it suffices to look for those that are of degree 1. For the case in which the space is 4-dimensional we have the following

**Theorem:** In a space of 4-dimensions any scalar-tensor Lagrangian which provides second-order

Euler-Lagrange tensor densities that are biconformally invariant of degree 1 is equivalent to the Lagrangian

$$L_{C,k=1} := g^{\frac{1}{2}}\kappa_4(\varphi^{-2}R + 6\varphi^{-4}X) + g^{\frac{1}{2}}\kappa_2\varphi^{-4} ,\qquad \text{Eq.1.26}$$

where $\kappa_2$ and $\kappa_4$ are constants with $X:=g^{ab}\varphi_{,a}\varphi_{,b}$.■

This theorem will be proved in the next section. As an immediate consequence of the **Theorem** and **Proposition 2** we have this

**Corollary:** In a space of 4-dimensions any scalar-tensor Lagrangian which provides second-order Euler-Lagrange tensor densities that are biconformally invariant of degree $k \neq 0$ is equivalent to the Lagrangian

$$L_{C,k} := g^{\frac{1}{2}}\kappa_4(\varphi^{-(2/k)}R + 6k^{-2}\varphi^{-(2/k+2)}X) + g^{\frac{1}{2}}\kappa_2\varphi^{-4/k} ,\qquad \text{Eq.1.27}$$

where $\kappa_2$ and $\kappa_4$ are real constants.■

I find the **Theorem** and its **Corollary** to be very interesting and analogous to a Theorem of Lovelock's, Ref. 7. Lovelock showed that in a space of 4-dimensions any Lagrangian which is a concomitant of a pseudo-Riemannian metric tensor and its derivatives of arbitrary order and which yields a second-order Euler-Lagrange tensor density must be equivalent to the Einstein Lagrangian with cosmological constant

$$L_E := \kappa g^{\frac{1}{2}} R + 2\Lambda g^{\frac{1}{2}}$$

where $\kappa$ and $\Lambda$ are constants. For the scalar-tensor case I have shown (*see,* Ref. 3, and also Deffayet *et al.* Ref.8, Kobyashi *et al.* Ref.9), that in a space of 4-dimensions any Lagrangain which is a concomitant of a pseudo-Riemannian metric tensor and scalar field and their derivatives of arbitrary order that yields second-order Euler-Lagrange tensor densities must be equivalent to

$$L_H = L_2 + L_3 + L_4 + L_5 \qquad \text{Eq.1.28}$$

where

$$L_2 := g^{\frac{1}{2}} G_2 \qquad \text{Eq.1.29}$$

$$L_3 := g^{½}G_3\Box\varphi \qquad \text{Eq.1.30}$$

$$L_4 := g^{½}G_4 R - 2g^{½}G_{4,X}((\Box\varphi)^2 - \varphi^{hi}\varphi_{hi}) \qquad \text{Eq.1.31}$$

and

$$L_5 := g^{½}G_5\varphi_{hi}G^{hi} + \tfrac{1}{3}g^{½}G_{5,X}((\Box\varphi)^3 - 3\Box\varphi\varphi^{hi}\varphi_{hi} + 2\varphi^h{}_i\varphi^i{}_j\varphi^j{}_h) \qquad \text{Eq.1.32}$$

where $G_2$, $G_3$, $G_4$ and $G_5$ are arbitrary real-vaued functions of $\varphi$ and X, with $_{,X}$ denoting a partial derivative with respect to X. The **Theorem** and its **Corollary** say that if we demand biconformal invariance of degree $k \neq 0$, then the four functions of $\varphi$ and X appearing in $L_H$ reduce to two constants with $L_3$ and $L_5$ falling by the wayside. In addition due to the work of Baker, *et al.* Ref. 10, P.Creminelli & F.Vernizzi, Ref. 11, J.Sakstein & B.Jain, Ref. 12, and J.M.Ezquiaga & M.Zumalaćarregui, Ref. 13, I can say that the field equations generated by the Lagrangians presented in Eqs.1.26 and 1.27 imply that the speed of light and gravity waves are equal for those scalar-tensor field theories.

I should like to point out that the **Corollary** tells us that the most general biconformally invariant second-order scalar-tensor field theory of degree $-1$ can be obtained from the Lagrangian

$$L_{C,k=-1} = g^{½}\kappa_4(\varphi^2 R + 6X) + g^{½}\kappa_2\varphi^4.$$

This Lagrangian has appeared numerous times and in various contexts in the literature. See, for example, S.Deser, Ref. 14, F.Englert, *et al.*, Ref. 15, F.Englert, *et al.* Ref. 16, and P.J.Steinhardt & N. Turok, Ref. 17.

To conclude this section I would like to make a few remarks about scalar-tensor field theories that involve multiple scalar fields.

Let $\alpha \in \mathbb{N}$ (the natural numbers) with $\alpha > 1$. I wish to consider scalar-tensor field theories in which there are $\alpha$ scalar fields, $\varphi_1, \ldots, \varphi_\alpha$. If L is the Lagrange scalar density for such a theory and a concomitant of $g_{ab}$, the $\varphi$'s and their derivatives of arbitrary, but finite order, then $\forall\, \beta = 1, \ldots, \alpha$, let $E_\beta(L)$ denote the Euler-Lagrange tensor density obtained by varying L with respect to $\varphi_\beta$, with the other fields held fixed. It can be shown, using an argument similar to the one employed to establish Eq.1.3, that the Euler-Lagrange tensor densities of L are related by

$$E_a{}^b(L)_{|b} + \tfrac{1}{2}(\varphi_{1,a}E_1(L) + \ldots + \varphi_{\alpha,a}E_\alpha(L)) = 0 \,.$$

If $k_1,\ldots,k_\alpha \in \mathbb{R}$, I define a multiconformal transformation of degree $(k_1,\ldots,k_\alpha)$ by $(g_{ab}, \varphi_1,\ldots,\varphi_\alpha) \to (g'_{ab}, \varphi'_1,\ldots,\varphi'_\alpha)$ where $g'_{ab} := e^{2\sigma}g_{ab}$, $\varphi'_1 := \exp(k_1\sigma)\varphi_1,\ldots, \varphi'_\alpha := \exp(k_\alpha\sigma)\varphi_\alpha$ where $\sigma$ is an arbitrary scalar field. If L is a multiscalar Lagrangian then it is said to be multiconformally invariant of degree $(k_1,\ldots,k_\alpha)$ up to a divergence if it satisfies the obvious generalization of Eq.1.7. Using an argument similar to the ones we used to prove **Propositions 1** and **2** we can easily establish the following propositions:

**Proposition 3:** Let $E^{ab}(L)$, and $E_\beta(L)$, $\beta=1,\ldots,\alpha$ be the Euler-Lagrange tensor densities of a multiscalar-tensor field theory. This theory will be multiconformally invariant of degree $(k_1,\ldots,k_\alpha)$ if and only if

$$2g_{ab}E^{ab}(L) + k_1\varphi_1 E_1(L) + \ldots + k_\alpha\varphi_\alpha E_\alpha(L) = 0. \qquad \text{Eq.1.33}$$

If $E^{ab}(L)$ and $E_\beta(L)$ satisfy Eq.1.33 then L is multiconformally invariant of degree $(k_1,\ldots,k_\alpha)$ up to a divergence.∎

**Proposition 4:** Let L be the Lagrange scalar density of a multiconformally invariant scalar-tensor field theory of degree $(1,\ldots,1)$. If $k_1,\ldots,k_\alpha \in \mathbb{R}\setminus\{0\}$, let $L(k_1,\ldots,k_\alpha)$ be the Lagrange scalar density concomitant of $g_{ab}, \varphi_1,\ldots,\varphi_\alpha$ obtained from L by replacing $\varphi_1,\ldots,\varphi_\alpha$ and their derivatives everywhere in L by $\varphi_1^{1/k_1},\ldots, \varphi_\alpha^{1/k_\alpha}$. Then $L(k_1,\ldots,k_\alpha)$ is the Lagrangian of a multiconformally invariant scalar-tensor field theory of degree $(k_1,\ldots,k_\alpha)$.∎

**Section 2: Proof of the Theorem**

The basic idea behind the proof of the **Theorem** is quite simple. Let L be a scalar-tensor

Lagrangian in a 4-dimensional space that yields second-order Euler-Lagrange tensor densities. I let

$$V(L) := 2g_{ab}E^{ab}(L) + \varphi E(L) .  \quad\quad\quad\quad \text{Eq.2.1}$$

Note that if we evaluate V(L) for the conformal variation $g(t)_{ab}$, $\varphi(t)$ given in Eq.1.18, then we would get the variation of L (with k=1) up to a factor of σ, as shown in Eq.1.22. So V(L) is essentially the (infinitesimal) conformal variation of L, and due to **Proposition 1**, we need it to vanish if L is to provide a biconformally invariant scalar-tensor field theory of degree 1. Due to Refs. 2 & 3, L must be equivalent to one of the $L_H$ Lagrangians given in Eq.1.28. So I shall assume that L is of second-order. Now a cursory examination of the Euler-Lagrange tensors associated with $L_2$, $L_3$, $L_4$ and $L_5$ (*see*, Eqs.4.12-4.17 in Ref. 3, along with the term in curly brackets on the right-hand side of Eq.3.21, which is essentially $E(L_H)$) shows that in general V(L) is a combination of terms involving $R^{abcd}R_{pqrs}$, $R^{abcd}\varphi_{ef}$, $R^{abcd}$, $\varphi^{ab}\varphi^{cd}\varphi^{ef}$, $\varphi^{ab}\varphi^{cd}$, $\varphi^{ab}$ and 1 (with various contractions involving $g_{pq}$ and $\varphi_r$), multiplied by coefficient functions of φ and X built from the G's and their partial derivatives. These terms are algebrically independent. Thus if V(L)=0, then each of these coefficient functions must vanish. This in turn leads to partial differential equations for the G's. Fortunately these partial differential equations are very easy to solve. Once we find all of the G's the **Theorem** will be proved. Now for the details.

It would be folly to begin by simply writing out $V(L_H)$ =0, and then trying to find the partial differential equations governing the G's. I found that it is easier to start with $V(L_5)$, and then progressively add $V(L_4)$, $V(L_3)$ and $V(L_2)$ to the mix, using what we discover about the G's at each stage to simplify $V(L_H)$.

To begin for $L_5$ given in Eq.1.32 we can use Eq.4.15 in Ref.3 (noting that $M_4$ in Refs. 2 and 3 is our $-¼G_5$, and that the Euler-Lagrange tensors in Refs. 2 and 3 are the negative of the ones used here) to find

$$g_{ab}E^{ab}(L_5) = g^{\frac{1}{2}}\{-G_{5X}X\varphi_{ab}G^{ab} - G_{5\varphi}\varphi_a\varphi_b R^{ab} - \tfrac{1}{3}(XG_{5XX}+G_{5X})[(\Box\varphi)^3 - 3\Box\varphi\varphi^{ab}\varphi_{ab} + 2\varphi^a{}_b\varphi^b{}_c\varphi^c{}_a] +$$

$$+ G_{5\varphi}((\Box\varphi)^2 - \varphi^{ab}\varphi_{ab}) + 2G_{5\varphi X}[\Box\varphi\varphi^a\varphi^b\varphi_{ab} - \varphi^a\varphi^b\varphi_{ac}\varphi_b{}^c] + G_{5\varphi\varphi}(X\Box\varphi - \varphi^a\varphi^b\varphi_{ab})\}. \qquad \text{Eq.2.2}$$

Unfortunately I did not compute $E(L_\alpha)$, for $\alpha=2,3,4,5$ explicitly in Ref.3, although these tensors are cryptically contained within the term in curly brackets on the right-hand side of Eq.3.21 in Ref.3. So these quantities will need to be computed here. To that end in Ref.2 I derive an expression for $E(L)$ when $L$ is a second-order Lagrangian. This expression is given by:

$$E(L) = \Phi - \Phi^a{}_{|a} + \Phi^{ab}{}_{|ab} \qquad \text{Eq.2.3}$$

with

$$\Phi := \frac{\partial L}{\partial \varphi}; \quad \Phi^a := \frac{\partial L}{\partial \varphi_{,a}} + \Gamma^a{}_{bc}\Phi^{bc} \quad \text{and} \quad \Phi^{ab} := \frac{\partial L}{\partial \varphi_{,ab}} \qquad \text{Eq.2.4}$$

(*see*, Eq.4.10 in Ref.3, where I use different symbols in $E(L)$). $\Phi$, $\Phi^a$ and $\Phi^{ab}$ are tensor densities known as the tensorial derivatives of L with respect to $\varphi$, $\varphi_{,a}$ and $\varphi_{,ab}$, or collectively, as the scalar tensorial derivatives of L. Computing $\Phi^a$ is actually easier than it looks, since to do so all one needs to do is compute the partial derivative of L with respect to $\varphi_{,a}$ regarding $\varphi$, $\varphi_{,a}$ and $\varphi_{,ab}$ as independent variables in L. In this way we find that the three scalar tensorial derivatives of $L_5$ are:

$$\Phi_5 = g^{\frac{1}{2}}\{G_{5\varphi}\varphi_{hi}G^{hi} + \tfrac{1}{3}G_{5\varphi X}[(\Box\varphi)^3 - 3\Box\varphi\varphi^{hi}\varphi_{hi} + 2\varphi^h{}_i\varphi^i{}_j\varphi^j{}_h]\} \qquad \text{Eq.2.5}$$

$$\Phi_5{}^a = g^{\frac{1}{2}}\{2G_{5X}\varphi^a\varphi_{hi}G^{hi} + \tfrac{2}{3}G_{5XX}\varphi^a[(\Box\varphi)^3 - 3\Box\varphi\varphi^{hi}\varphi_{hi} + 2\varphi^h{}_i\varphi^i{}_j\varphi^j{}_h]\} \qquad \text{Eq.2.6}$$

$$\Phi_5{}^{ab} = g^{\frac{1}{2}}\{G_5 G^{ab} + G_{5X}[g^{ab}(\Box\varphi)^2 - g^{ab}\varphi^{hi}\varphi_{hi} - 2\Box\varphi\varphi^{ab} + 2\varphi^{ah}\varphi^b{}_h]\}. \qquad \text{Eq.2.7}$$

It is an exeptionally tedious task to compute $E(L_5)$ using Eq.2.3, and I leave that to you. What you will find, among other things, is that the only the term quadratic in the curvature tensor is

$$g^{\frac{1}{2}}G_{5X}[2\varphi^a\varphi^b R_{ac}R_b{}^c - \varphi^a\varphi^b R_{acde}R_b{}^{cde} + 2\varphi^a\varphi^b G^{cd}R_{acbd}].$$

This term can be rewritten using the fact that in a 4-dimensional space (*see,* Lanczos Ref.18 and Lovelock Ref. 19)

$$0 = \delta^{acdef}{}_{bhijk}\varphi_a\varphi^b R_{cd}{}^{hi}R_{ef}{}^{jk}$$

where $\delta^{acdef}_{bhijk}$ is the 5x5 generalized Kronecker delta. This equation implies that

$$2\varphi^a\varphi^b G^{cd}R_{acbd} + 2\varphi^a\varphi^b R_{ac}R_b{}^c - \varphi^a\varphi^b R_{acde}R_b{}^{cde} = \tfrac{1}{4}X[R^2 - 4R_{ab}R^{ab} + R_{abcd}R^{abcd}] \ .$$

Thus the ony term in $V(L_5)$ which is quadratic in the curvature tensor is

$$\tfrac{1}{4}g^{1/2}G_{5X}X[R^2 - 4R_{ab}R^{ab} + R_{abcd}R^{abcd}] \ ,$$

and hence

$$G_{5X} = 0 \qquad\qquad\qquad\qquad\qquad\qquad\qquad\qquad\qquad\qquad \text{Eq.2.8}$$

to guarantee that $V(L_5) = 0$. In fact since $V(L_2)$, $V(L_3)$ and $V(L_4)$ are independent of quadratic curvature terms we can deduce that Eq.2.8 must hold if $V(L_H) = 0$.

Eq.2.8 causes many of the terms in $E(L_5)$ to vanish, thereby enabling us to use Eqs.2.2-2.7 to show that

$$V(L_5) = g^{1/2}\{-2G_{5\varphi}\varphi^a\varphi^b R_{ab} + 2G_{5\varphi}((\Box\varphi)^2 - \varphi^{ab}\varphi_{ab}) + 2G_{5\varphi\varphi}(X\Box\varphi - \varphi^a\varphi^b\varphi_{ab}) +$$

$$+ G_{5\varphi\varphi}\varphi\varphi^a\varphi^b G_{ab} + 2\varphi G_{5\varphi}\varphi_{ab}G^{ab}\} \ . \qquad\qquad \text{Eq.2.9}$$

If we want $L_5$ to be biconformally invariant of degree 1, then the term involving $\varphi_{ab}G^{ab}$ in Eq.2.9 would imply that $G_{5\varphi}$ must vanish. Combining this with Eq.2.8 shows us that $L_5$ is biconformally invariant of degree 1 if and only if $G_5$ is a constant, in which case $L_5$ is a divergence and its Euler-Lagrange tensors vanish. Hence $L_5$ can not generate useful biconformally invariant field equations.

We shall now explore $V(L_4+L_5)$ to see if it might be possible to obtain biconformaly invariant field tensor densities from $L_4+L_5$.

Upon noticing that the Lagrangian $L_2$ in Eq.4.3 of Ref. 3 is our Lagrangian $L_4$ when the function $M_2$ in Ref.3 is replaced by $\tfrac{1}{2}G_4$, we can then use Eq.4.13 of Ref.3 to deduce that (recall the Euler-Lagrange tensor densities in Ref.3 are the negative of ours)

$$g_{ab}E^{ab}(L_4) = g^{1/2}\{-3G_{4\varphi\varphi}X - 3G_{4\varphi}\Box\varphi + 2G_{4\varphi X}[X\Box\varphi - 4\varphi^a\varphi^b\varphi_{ab}] +$$

$$+ 2G_{4XX}[2\varphi^a\varphi^b\varphi_{ab}\Box\varphi - 2\varphi^a\varphi^b\varphi_{ac}\varphi_b{}^c + X((\Box\varphi)^2 - \varphi^{ab}\varphi_{ab})] +$$

$$+ 2G_{4X}[(\Box\varphi)^2 - \varphi^{ab}\varphi_{ab}] + G_4 R - G_{4X}[XR + 2\varphi^a\varphi^b R_{ab}]\} . \qquad \text{Eq.2.10}$$

To compute $E(L_4)$ we need the scalar tensorial derivatives of $L_4$, and they are given by

$$\Phi_4 = g^{\frac{1}{2}}\{G_{4\varphi}R - 2G_{4\varphi X}[(\Box\varphi)^2 - \varphi^{hi}\varphi_{hi}]\}$$

$$\Phi_4^a = 2g^{\frac{1}{2}}G_{4X}\varphi^a R - 4g^{\frac{1}{2}}G_{4XX}\varphi^a[(\Box\varphi)^2 - \varphi^{hi}\varphi_{hi}]$$

$$\Phi_4^{ab} = 4g^{\frac{1}{2}}G_{4X}(\varphi^{ab} - \Box\varphi g^{ab}) .$$

Using these in Eq.2.3 we find, with some effort, that

$$E(L_4) = g^{\frac{1}{2}}\{8G_{4XX}[2\varphi^c\varphi_{ca}\varphi_b R^{ab} + \varphi^a\varphi^b\varphi^{cd}R_{acbd} - \tfrac{1}{2}\varphi^a\varphi^b\varphi_{ab}R - \varphi^a\varphi^b\Box\varphi R_{ab}] + 8G_{4\varphi X}\varphi^a\varphi^b R_{ab} +$$

$$- 2G_{4\varphi X}XR + 4G_{4X}\varphi^{ab}G_{ab} + G_{4\varphi}R + 8G_{4XXX}[2\varphi^a\varphi^b\varphi_{ac}\varphi_{bd}\varphi^{cd} - 2\varphi^a\varphi^b\varphi_{ac}\varphi_b{}^c\Box\varphi +$$

$$+ \varphi^a\varphi^b\varphi_{ab}(\Box\varphi)^2 - \varphi^a\varphi^b\varphi_{ab}\varphi^{cd}\varphi_{cd}] + 4G_{4XX}[(\Box\varphi)^3 - 3\varphi^{ab}\varphi_{ab}\Box\varphi + 2\varphi^a{}_b\varphi^b{}_c\varphi^c{}_a] +$$

$$+ 4G_{4\varphi XX}[4\varphi^a\varphi^b\varphi_{ac}\varphi_b{}^c - 4\varphi^a\varphi^b\varphi_{ab}\Box\varphi + X(\Box\varphi)^2 - X\varphi^{ab}\varphi_{ab}] + 4G_{4\varphi\varphi X}[\varphi^a\varphi^b\varphi_{ab} - X\Box\varphi] +$$

$$+ 6G_{4\varphi X}[\varphi^{ab}\varphi_{ab} - (\Box\varphi)^2]\} . \qquad \text{Eq.2.11}$$

Upon examining Eqs.2.10 and 2.11 we see that $g_{ab}E^{ab}(L_4)$ does not contain a term involving the curvature times $\varphi_{ab}$ while $E(L_4)$ does. Hence if $V(L_4) = 0$, then we must have this term vanish, giving us

$$8G_{4XX}[2\varphi^c\varphi_{ca}\varphi_b R^{ab} + \varphi^a\varphi^b\varphi^{cd}R_{acbd} - \tfrac{1}{2}\varphi^a\varphi^b\varphi_{ab}R - \varphi^a\varphi^b\Box\varphi R_{ab}] + 4G_{4X}\varphi^{ab}G_{ab} = 0. \qquad \text{Eq.2.12}$$

This must hold for all scalar fields and metric tensors. In particular it must hold for those with $R_{abcd} \neq 0$ and $G_{ab} = 0$. The choice of a metric with $G_{ab}=0$, does not effect the possible values of $G_4$, $G_{4X}$ and $G_{4XX}$. Thus Eq.2.12 first gives us $G_{4XX} = 0$, and then $G_{4X} = 0$, when $V(L_4) = 0$. If we now combine this fact with Eqs.2.10 and 2.11 we see that $V(L_4) = 0$ requires

$$-6G_{4\varphi\varphi}X - 6G_{4\varphi}\Box\varphi + 2G_4 R + \varphi G_{4\varphi}R = 0 . \qquad \text{Eq.2.13}$$

The $\Box\varphi$ term in Eq.2.13 tells us that $G_{4\varphi} = 0$, and using this back in Eq.2.13 shows us that $G_4 = 0$. Hence if $V(L_4) = 0$, then $L_4$ vanishes.

So right now we know that neither $L_4$ nor $L_5$ can provided biconformally invariant scalar-tensor theories of degree 1. Let us see what happens when we consider $L_4 + L_5$. Since V is a linear operator we can use Eqs.2.9-2.11 to deduce that the term involving $R^{abcd}\varphi_{ef}$ in the equation $V(L_4+L_5)=0$ implies that

$$2\varphi G_{5\varphi}\varphi_{ab}G^{ab} + 4\varphi G_{4XX}[4\varphi^c\varphi_{ca}\varphi_b R^{ab} + 2\varphi^a\varphi^b\varphi^{cd}R_{acbd} - \varphi^a\varphi^b\varphi_{ab}R - 2\varphi^a\varphi^b\Box\varphi R_{ab}] +$$
$$+ 4\varphi G_{4X}\varphi^{ab}G_{ab} = 0 .$$
Eq.2.14

As I mentioned earlier this equation must hold for every metric tensor and scalar field, in particular for those with $R^{abcd} \neq 0$, and $G_{ab} = 0$. Hence Eq.2.14 tells us that

$$G_{4XX} = 0 \text{ and } G_{5\varphi} = -2G_{4X} .$$
Eq.2.15

Since there does not exist a term in $V(L_3)$ and $V(L_2)$ involving $R^{abcd}\varphi_{ef}$ we can conclude that Eq.2.15 must hold if $V(L_H) = 0$. So at present we can employ Eqs.2.9-2.11 to write

$$V(L_4+L_5) = g^{\frac{1}{2}}\{6\varphi G_{4\varphi X}\varphi^a\varphi^b R_{ab} + R[2G_4 + \varphi G_{4\varphi} - 2XG_{4X} - \varphi XG_{4\varphi X}] + 6\varphi G_{4\varphi X}[\varphi^{ab}\varphi_{ab} - (\Box\varphi)^2] +$$
$$- \Box\varphi(6G_{4\varphi} + 4\varphi XG_{4\varphi\varphi X}) + 4\varphi^a\varphi^b\varphi_{ab}[\varphi G_{4\varphi\varphi X} - 2G_{4\varphi X}] - 6XG_{4\varphi\varphi}\} .$$
Eq.2.16

Thus if we want $V(L_4+L_5) = 0$ we can use Eq.2.16 to conclude that $G_{4\varphi X} = 0$, due to the term involving $\varphi^a\varphi^b R_{ab}$, and then that $G_{4\varphi} = 0$, due to the $\Box\varphi$ term. Hence the term involving R in Eq.2.16 now permits us to deduce that $G_4 = \kappa X$, where $\kappa$ is a constant. This can be combined with Eq.2.15 to give us $G_5 = -2\kappa\varphi$, and consequently Eqs.1.31 and 1.32 imply that

$$L_4 + L_5 = g^{\frac{1}{2}}\{-2\kappa\varphi\varphi_{ab}G^{ab} + \kappa XR - 2\kappa((\Box\varphi)^2 - \varphi^{ab}\varphi_{ab})\} .$$
Eq.2.17

Thus at first sight it appears that we have a biconformally invariant scalar-tensor field theory of degree 1. But it is easy to show that the Lagrangian given in Eq.2.17 can be rewritten as

$$L_4 + L_5 = - [2\kappa g^{\frac{1}{2}}(\varphi\varphi_b G^{ab} + \Box\varphi\varphi^a - \varphi^{ab}\varphi_b)]_{|a} .$$

Consequently the Lagrangian presented in Eq.2.17 yields trivial field equations, and so we have shown that $L_4+L_5$ is incapable of producing second-order biconformally invariant scalar-tensor field equations of degree 1. So let us now add $L_3$ to the mix to see if we fair any better in our quest for biconformally invariant scalar-tensor theories.

Using Eq.4.12 in Ref.3 with $M_1$ replaced by $G_3$ we obtain

$$g_{ab}E^{ab}(L_3) = -g^{\frac{1}{2}}\{XG_{3X}\Box\varphi + 2G_{3X}\varphi^a\varphi^b\varphi_{ab} + XG_{3\varphi}\}. \qquad \text{Eq.2.18}$$

To derive $E(L_3)$ we require the scalar tensorial derivatives of $L_3$, which using Eq.1.30 are easily found to be

$$\Phi_3 = g^{\frac{1}{2}}G_{3\varphi}\Box\varphi, \quad \Phi_3{}^a = 2g^{\frac{1}{2}}G_{3X}\varphi^a\Box\varphi \quad \text{and} \quad \Phi_3{}^{ab} = g^{\frac{1}{2}}G_3 g^{ab}.$$

Thus we can use Eq.2.3 to deduce that

$$E(L_3) = g^{\frac{1}{2}}\{2G_{3X}\varphi^a\varphi^b R_{ab} + 2G_{3X}\varphi^{ab}\varphi_{ab} - 2G_{3X}(\Box\varphi)^2 + 4G_{3XX}\varphi^a\varphi^b\varphi_{ac}\varphi_b{}^c - 4G_{3XX}\varphi^a\varphi^b\varphi_{ab}\Box\varphi +$$
$$+ 2[G_{3\varphi} - G_{3\varphi X}X]\Box\varphi + 4G_{3\varphi X}\varphi^a\varphi^b\varphi_{ab} + XG_{3\varphi\varphi}\}. \qquad \text{Eq.2.19}$$

Employing Eqs.2.18 and 2.19 it is easy to show that $V(L_3)=0$ if and only if $G_3$ is a constant. In that case $L_3$ is a divergence and hence its Euler-Lgrange tensor densities vanish. Consequently $L_3$ is incapable of generating a biconformally invariant scalar-tensor field theory.

So far we have demonstrated that if $V(L_H)=0$ then we must have

$$G_{5X} = 0, \; G_{4XX} = 0 \text{ and } G_{5\varphi} = -2G_{4X}. \qquad \text{Eq.2.20}$$

Thus we can employ Eqs.2.16, 2.18 and 2.19 to conclude that if $V(L_H)=0$, then

$$V(L_3+L_4+L_5) = g^{\frac{1}{2}}\{\varphi^a\varphi^b R_{ab}[6\varphi G_{4\varphi X} + 2\varphi G_{3X}] + R[2G_4 + \varphi G_{4\varphi} - 2XG_{4X} - \varphi XG_{4\varphi X}] +$$
$$- 4\varphi G_{3XX}\varphi^a\varphi^b\varphi_{ab}\Box\varphi + \varphi^{ab}\varphi_{ab}[6\varphi G_{4\varphi X} + 2\varphi G_{3X}] - (\Box\varphi)^2[6\varphi G_{4\varphi X} + 2\varphi G_{3X}] +$$
$$+ 4\varphi G_{3XX}\varphi^a\varphi^b\varphi_{ac}\varphi_b{}^c - \Box\varphi[6G_{4\varphi} + 4\varphi XG_{4\varphi\varphi X} + 2XG_{3X} - 2\varphi G_{3\varphi} + 2\varphi XG_{3\varphi X}] +$$
$$+ 4\varphi^a\varphi^b\varphi_{ab}[\varphi G_{4\varphi\varphi X} - 2G_{4\varphi X} - G_{3X} + \varphi G_{3\varphi X}] - 6XG_{4\varphi\varphi} - 2XG_{3\varphi} + \varphi XG_{3\varphi\varphi}\}. \qquad \text{Eq.2.21}$$

From Eq.2.21 we immediately see that if

$$V(L_3 + L_4 + L_5) = 0, \qquad \text{Eq.2.22}$$

then

$$G_{3X} = -3G_{4\varphi X} \qquad \text{Eq.2.23}$$

and

$$2G_4 + \varphi G_{4\varphi} - 2XG_{4X} - \varphi XG_{4\varphi X} = 0. \qquad \text{Eq.2.24}$$

It is fairly easy to demonstrate that the set

$$\{(\Box\varphi)^2, \varphi^{ab}\varphi_{ab}, \varphi^a\varphi^b\varphi_{ab}\Box\varphi, \varphi^a\varphi^b\varphi_{ac}\varphi_b{}^c\}$$

is linearly independent in a 4-dimensional space over the ring of real valued functions of $\varphi$ and X. To do so take a linear combination of these terms and set it equal to zero. Differentiate that expression with respect to $\varphi_{pq}$ and $\varphi_{rs}$ to obtain a four index contravariant tensor field that is identically equal to zero in a 4-dimensional space. To prove that each term in that expression must vanish just look at various contractions of that tensor with $\varphi_u$, $\xi_v$ and $\zeta_w$ where the latter two covariant vector fields are perpendicular to $\varphi_u$ and each other. Hence we can now use Eq.2.21 to deduce that when Eq.2.22 holds

$$G_{3XX} = 0 . \qquad \text{Eq.2.25}$$

Since $V(L_2)$ is algebraically at most of first degree in the second derivatives of $\varphi$, we can conclude that Eqs.2.20 and 2.25 must hold if $V(L_H) = 0$. So we can now write

$$G_5 = \mu_5, \ G_4 = \mu_4 X + \nu_4 \ \text{and} \ G_3 = \mu_3 X + \nu_3 \qquad \text{Eq.2.26}$$

where the $\mu$'s and $\nu$'s are differentiable functions of $\varphi$. Eq.2.26 must also hold when $V(L_H)=0$.

Due to Eqs.2.20, 2.23 and 2.24 we have

$$\mu_5' = -2\mu_4 \ \text{and} \ \mu_3 = -3\mu_4', \qquad \text{Eq.2.27}$$

where here a ' denotes a derivative with respect to $\varphi$, and not a conformal transformation.

Using the expression involving $G_4$ in Eq.2.24 we discover that

$$\nu_4 = \kappa_4 \varphi^{-2} \qquad \text{Eq.2.28}$$

where $\kappa_4$ is constant.

Thus far what we have done also applies to determining the values of $G_3$, $G_4$ and $G_5$ in $V(L_H)=0$. We shall now try to determine the values of these functions exclusively for $V(L_3+L_4+L_5) = 0$. To that end since the set $\{\Box\varphi, \varphi^a\varphi^b\varphi_{ab}, 1\}$ is linearly independent over the ring of real valued functions of φ and X, Eq.2.21 generates the following restrictions:

$$3G_{4\varphi} + 2\varphi X G_{4\varphi\varphi X} + X G_{3X} - \varphi G_{3\varphi} + \varphi X G_{3\varphi X} = 0 \qquad \text{Eq.2.29}$$

$$\varphi G_{4\varphi\varphi X} - 2G_{4\varphi X} - G_{3X} + \varphi G_{3\varphi X} = 0 \qquad \text{Eq.2.30}$$

$$6G_{4\varphi\varphi} + 2G_{3\varphi} - \varphi G_{3\varphi\varphi} = 0 \, . \qquad \text{Eq.2.31}$$

At present owing to Eqs.2.26-2.28 we know that

$$G_4 = \mu_4 X + \kappa_4 \varphi^{-2} \quad \text{and} \quad G_3 = -3\mu_4' X + \nu_3 \, . \qquad \text{Eq.2.32}$$

If we use this in Eq.2.31 we find that $\mu_4$ and $\nu_3$ must satisfy

$$\mu_4''' = 0 \, , \; 36\kappa_4 \varphi^{-4} + 2\nu_3' - \varphi \nu_3'' = 0$$

and hence

$$\mu_4 = \mu_{41}\varphi^2 + \mu_{42}\varphi + \mu_{43} \text{ and } \nu_3 = \tfrac{1}{3}\nu_{31}\varphi^3 + 2\kappa_4 \varphi^{-3} + \nu_{32} \qquad \text{Eq.2.33}$$

where $\kappa_4$, $\mu_{41}$, $\mu_{42}$, $\mu_{43}$, $\nu_{31}$ and $\nu_{32}$ are constants. Using Eqs.2.32 and 2.33 in Eq.2.30 gives us

$$\mu_{41} = \mu_{42} = 0. \qquad \text{Eq.2.34}$$

Consequently we can use Eq.2.29 to deduce that

$$\nu_{31} = 0. \qquad \text{Eq.2.35}$$

Upon combining Eqs 2.32-2.35 we find

$$G_3 = 2\kappa_4 \varphi^{-3} + \nu_{32} \text{ and } G_4 = \mu_{43} X + \kappa_4 \varphi^{-2} \, . \qquad \text{Eq.2.36}$$

Since Eq.2.20 tells us that $G_{5\varphi} = -2G_{4X}$ we can employ Eq.2.36 to obtain

$$G_5 = -2\mu_{43}\varphi + \mu_{51} \qquad \text{Eq.2.37}$$

where $\mu_{51}$ is a constant. So owing to Eqs.1.30-1.32, 2.36 and 2.37 we can write

$$L_3 + L_4 + L_5 = g^{1/2}\{[-2\mu_{43}\varphi + \mu_{51}]\varphi_{ab}G^{ab} + \mu_{43}XR + \kappa_4\varphi^{-2}R - 2\mu_{43}[(\Box\varphi)^2 - \varphi^{ab}\varphi_{ab}] +$$

$$+ 2\kappa_4\varphi^{-3}\Box\varphi + \nu_{32}\Box\varphi\} \, . \qquad \text{Eq.2.38}$$

The terms involving $\mu_{51}$ and $\nu_{32}$ can be immediately dismissed since they are divergences. Likewise the terms involving $\mu_{43}$ are a divergence due to Eq.2.17 (with $\kappa=\mu_{43}$) and the remarks following it.

Thus modulo divergences

$$L_3 + L_4 + L_5 = g^{\frac{1}{2}}(\kappa_4\varphi^{-2}R + 2\kappa_4\varphi^{-3}\Box\varphi) \,. \qquad \text{Eq.2.39}$$

The Lagrangian presented in Eq.2.39 yields a second-order biconformally invariant scalar-tensor field theory of degree 1. Interestingly enough this Lagrangian is actually of the $L_2 + L_4$ type since

$$2g^{\frac{1}{2}}\kappa_4\varphi^{-3}\Box\varphi = (2g^{\frac{1}{2}}\kappa_4\varphi^{-3}\varphi^a)_{|a} + 6g^{\frac{1}{2}}\kappa_4\varphi^{-4}X \,.$$

Thus the most general second-order biconformally invariant scalar-tensor field theory of degree 1 derivable from a Lagrangian of the $L_3 + L_4 + L_5$ can be obtained from either Eq.2.39, or equivalently

$$L_{2,4} := g^{\frac{1}{2}}(\kappa_4\varphi^{-2}R + 6\kappa_4\varphi^{-4}X) \,. \qquad \text{Eq.2.40}$$

I shall now complete the proof of the **Theorem** by adjoining the Lagrangian $L_2$ to the above work.

Using Eq.4.17 in Ref. 3 along with Eq.2.3 it is easy to show that

$$V(L_2) = g^{\frac{1}{2}}\{-2\varphi G_{2X}\Box\varphi - 4\varphi G_{2XX}\varphi^a\varphi^b\varphi_{ab} + 4G_2 + \varphi G_{2\varphi} - 2XG_{2X} - 2\varphi XG_{2\varphi X}\} \,. \qquad \text{Eq.2.41}$$

If we require $V(L_2) = 0$, then we find that

$$L_2 = g^{\frac{1}{2}}\kappa_2\varphi^{-4} \,, \qquad \text{Eq.2.42}$$

where $\kappa_2$ is a constant. Hence of the four Lagrangians $L_2$, $L_3$, $L_4$ and $L_5$ only $L_2$ will yield a biconformally invariant second-order scalar-tensor field theory of degree 1 without the help of other Lagrangians.

Let us next examine $L_H$ to see what results. From Eqs. 2.21 and 2.41 we see that the inclusion of $L_2$ into our analysis only effects the terms involving $\Box\varphi$, $\varphi^a\varphi^b\varphi_{ab}$ and 1 (when working over the ring of real valued functions of $\varphi$ and X). This observation leads to the following three equations

$$6G_{4\varphi} + 4\varphi XG_{4\varphi\varphi X} + 2XG_{3X} - 2\varphi G_{3\varphi} + 2\varphi XG_{3\varphi X} + 2\varphi G_{2X} = 0 \qquad \text{Eq.2.43}$$

$$\varphi G_{4\varphi\varphi X} - 2G_{4\varphi X} - G_{3X} + \varphi G_{3\varphi X} - \varphi G_{2XX} = 0 \qquad \text{Eq.2.44}$$

$$6XG_{4\varphi\varphi} + 2XG_{3\varphi} - \varphi XG_{3\varphi\varphi} - 4G_2 + 2XG_{2X} - \varphi G_{2\varphi} + 2\varphi XG_{2\varphi X} = 0. \qquad \text{Eq.2.45}$$

From our previous analysis of those terms in $V(L_3+L_4+L_5)$, which did not involve $\Box\varphi$, $\varphi^a\varphi^b\varphi_{ab}$ and 1, we know that

$$G_{5X}=0, \ G_{5\varphi} = -2G_{4X}, \ G_{3X} = -3G_{4\varphi X}, \ G_{4XX} = 0 \text{ and } G_{3XX} = 0. \qquad \text{Eq.2.46}$$

That analysis enabled us to write (*see*, Eqs.2.26 and 2.32)

$$G_5 = \mu_5, \ G_4 = \mu_4 X + \kappa_4\varphi^{-2} \text{ and } G_3 = -3\mu_4'X + \nu_3 \qquad \text{Eq.2.47}$$

where Eq.2.46 tells us that $\mu_5' = -2\mu_4$. Combining Eqs.2.43 and 2.47 gives us

$$-24\kappa_4\varphi^{-3} + 4\varphi X\mu_4'' - 2\varphi\nu_3' + 2\varphi G_{2X} = 0. \qquad \text{Eq.2.48}$$

Upon differentiating this equation with respect to X twice we find that $G_{2XXX} = 0$, and hence

$$G_2 = \mu_{21}X^2 + \mu_{22}X + \nu_2$$

where $\mu_{21}$, $\mu_{22}$ and $\nu_2$ are constants. Using this expression back in Eq.2.48 shows that

$$\mu_{21} = -\mu_4'' \text{ and } \mu_{22} = \nu_3' + 6\kappa_4\varphi^{-4},$$

and hence

$$G_2 = -\mu_4''X^2 + (\nu_3' + 6\kappa_4\varphi^{-4})X + \nu_2. \qquad \text{Eq.2.49}$$

We now direct our attention to Eq.2.44. The expressions for $G_2$, $G_3$ and $G_4$ given in Eqs.2.47 and 2.49 imply that $\mu_4' = 0$ and consequently $\mu_4$ is a constant. Thus Eqs.2.46, 2.47 and 2.49 tell us that

$$G_5 = \mu_4\varphi + \mu_{52}; \ G_4 = \mu_4 X + \kappa_4\varphi^{-2}; \ G_3 = \nu_3 \text{ and } G_2 = [\nu_3' + 6\kappa_4\varphi^{-4}]X + \nu_2 \qquad \text{Eq.2.50}$$

where $\mu_{52}$ is a constant with $\nu_2$ and $\nu_3$ still being functions of $\varphi$.

The last constraint equation at our disposal is Eq.2.45. Using the expressions for $G_2$, $G_3$ and $G_4$ from Eq.2.50 in Eq.2.45 shows that

$$\varphi\nu_2' + 4\nu_2 = 0,$$

and hence

$$\nu_2 = \kappa_2\varphi^{-4} \qquad \text{Eq.2.51}$$

where $\kappa_2$ is a constant.

So we can now combine Eqs.1.29-1.32, 2.50 and 2.51 to conclude that in a space of four-dimensions the most general second-order biconformally invariant scalar-tensor field theory of degree 1 obtainable from a variational principle can be obtained from the Lagrangian

$$\mathcal{L} := g^{½}\{[\mu_4\varphi + \nu_{51}]\varphi_{ab}G^{ab} + [\mu_4 X + \kappa_4\varphi^{-2}]R - 2\mu_4[(\Box\varphi)^2 - \varphi^{ab}\varphi_{ab}] + \nu_3\Box\varphi + \nu_3'X +$$
$$+ 6\kappa_4\varphi^{-4}X + \kappa_2\varphi^{-4}\} . \quad \text{Eq.2.52}$$

In Eq.2.52 the Lagrangian involving $\nu_{51}$ is a divergence, and the same is true for the Lagrangians involving $\mu_4$ and $\nu_3$. These three Lagrangians can be dropped from $\mathcal{L}$, leaving us with the Lagrangian $L_{C,k=1}$ of the **Theorem**. So the proof is now complete.

## Acknowledgement

I would like to thank Professor Paul J. Steinhardt for suggesting that I investigate conformal transformations of scalar-tensor field theories that affect both the metric tensor and scalar field.